\documentclass[conf]{IEEEtran}
\usepackage{blindtext, graphicx}
\usepackage{graphicx}
\usepackage{xcolor}
\usepackage{caption}
\usepackage{multirow}
\usepackage{placeins}

\usepackage[linesnumbered,ruled,vlined,lined,boxed,commentsnumbered]{algorithm2e}
\include{pythonlisting}
\usepackage{cite}

\usepackage[cmex10]{amsmath}
\usepackage{url}
\usepackage{tabto}
\usepackage{amsmath}

\begin{document}
\bstctlcite{IEEEexample:BSTcontrol}
    \title{Simulating Vehicle Movement and Multi-Hop Connectivity from Basic Safety Messages}
  \author{Noah~Carter, Mohammad~A.~Hoque, Md~Salman~Ahmed 
\\% <-this % stops a space
Department of Computing\\
   East Tennessee State University\\
   \{carterns, hoquem, ahmedm\}@etsu.edu
}

\author{\IEEEauthorblockN{Noah~Carter\IEEEauthorrefmark{1},
Mohammad~A.~Hoque\IEEEauthorrefmark{1},
Md~Salman~Ahmed\IEEEauthorrefmark{2}}\\
\IEEEauthorblockA{\IEEEauthorrefmark{1}Department of Computing, East Tennessee State University\\
\IEEEauthorrefmark{2}Department of Computer Science, Virginia Polytechnic Institute and State University\\
\IEEEauthorrefmark{1}\{carterns, hoquem\}@etsu.edu \\\IEEEauthorrefmark{2}ahmedms@vt.edu}}

% ====================================================================
\maketitle

\begin{abstract}
The Basic Safety Message (BSM) is a standardized communication packet that is sent every tenth of a second between connected vehicles using Dedicated Short Range Communication (DSRC). BSMs contain data about the sending vehicle's state, such as speed, location, and the status of the turn signal \cite{safety}. Presently, many BSM datasets from various United States locations are available through the connected vehicle testbeds of U.S. Department of Transportation. However, without a proper visualization tool, it is not possible to analyze or obtain a visual overview of the spatio-temporal distribution of the data. For this purpose, a web application has been developed which can ingest a raw BSM dataset and display a time-based simulation of vehicle movement. The simulation also displays multi-hop vehicular network connectivity for DSRC. This paper gives details about the application, including an explanation of the multi-hop partitioning algorithm used to classify the vehicles into separate network partitions. A performance analysis for the simulation is included, in which it is suggested that calculating a connectivity matrix with the multi-hop partitioning algorithm is computationally expensive for a large number of vehicles.
\end{abstract}

\IEEEpeerreviewmaketitle

% === I. INTRODUCTION =============================================================

\section{Introduction}

Dedicated Short Range Communication (DSRC) offers the potential for a variety of safety-critical applications that leverage vehicle-to-vehicle (V2V) and vehicle-to-infrastructure (V2I) communication modes. These applications follow standard communication protocols defined by SAE J2735 and IEEE 1609.x. One of the primary requirements for one class of safety applications is the broadcast of Basic Safety Messages (BSM) every tenth of a second through which the vehicles share their locations, speed, direction, and mobility information. When other vehicles and drivers have more information about the world around them, they can make better and safer decisions. This is the benefit of sending and receiving BSMs. Such communication is one example of Vehicle-to-Vehicle (V2V) communication, which has major implications for road safety and efficiency and is being widely studied.

DSRC is a short-range communication medium over which BSMs can be sent. However, because the range of DSRC is in many cases less than 1000 meters, it is necessary and useful to employ `multi-hop' communication. In multi-hop communication, messages traverse a network of vehicles (that act as message relays) in order to reach their destinations. Thus, rather than going directly from the source vehicle to the targets, the messages `hop' along an ad hoc network of closer vehicles to eventually reach as many targets as possible.

% === I. RELATED WORKS =============================================================
% =================================================================================
\section{Related Works}

\subsection{BSM Data Analysis}

Some BSM datasets are publicly available and are interesting subjects for data analysis. This includes, for example, the Safety Pilot Model Deployment project, in which tens of Michigan volunteers' vehicles were mounted with BSM-broadcasting devices. One aspect of such data analysis is visualization and simulation. Allowing an analyst to see the data in a helpful and intuitive format enables the analyst to make more informed and strategic decisions \cite{safety, Liu}. Our previous research has utilized BSM data for analyzing mobility patterns and developing various safety-critical applications in connected vehicle environments \cite{hoque2012analysis,elbery2015integrated,ahmed2016comparative,ahmed2016demo,ahmed2016partitioning,ahmed2017demo,jordan2017poster,ahmed2018intersection,Hoque,ITSMag}. 

The application described in this paper is intended to assist an analyst in visualizing location-related aspects of BSM data, specifically vehicle locations and vehicle connectivity partitions. To display the latter, it is necessary to calculate the partitions using a multi-hop partitioning algorithm.

\subsection{Multi-Hop Partitioning Algorithm}

Hoque et al. \cite{Hoque} introduced a multi-hop partitioning algorithm based on a modification of Warshall's algorithm. The purpose of the algorithm is to distinguish each vehicular ad hoc network partition--i.e., each connected component of vehicles that are within multi-hop communication range of each other. Again, vehicles can communicate directly, but they may also do so indirectly via one or more intermediate vehicles through which data and messages can pass (or `hop'). All vehicles that can communicate with each other by one of these two means are grouped together as a partition in the output of the algorithm. The result is a collection of partitions. The current work has further extended the algorithm developed by Hoque et. al. \cite{Hoque} to provide a graphical interface to visualize the connected components within a metro area using an empirically obtained BSM dataset.

\begin{figure}[h!]
\centering
\includegraphics{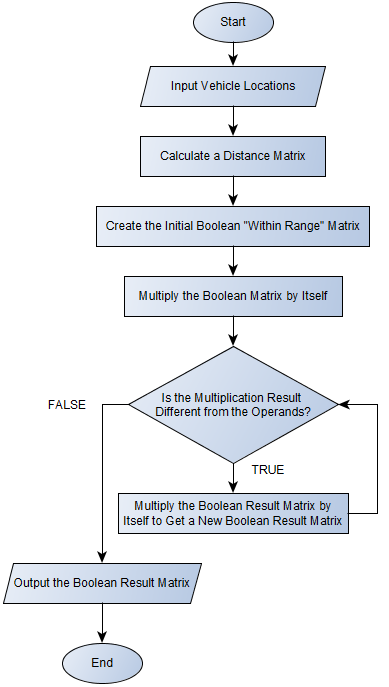}
\caption{Flowchart for a multi-hop partitioning algorithm, simplified from Hoque et al.'s \cite{Hoque} implementation.}
\label{multihopFlowchart}
\end{figure}

The algorithm begins (see Figure \ref{multihopFlowchart}) by reading in vehicle locations and calculating a square, symmetric `distance matrix.' Each element $a_{i,j}$ of the distance matrix is the calculated distance between the two vehicles that are represented by row $i$ and column $j$. Each entry of the distance matrix is then translated into a boolean value by determining whether it is less than or greater than the predetermined DSRC range (i.e., the maximum distance a given DSRC device can broadcast communications). Those distances that are less than the range become $1$; others become $0$. \textbf{Using boolean algebra,} the newly-formed boolean matrix is then multiplied by a copy of itself. This multiplication represents the first `hop,' and the result is a `connectivity matrix.'

The algorithm then proceeds with a series of boolean matrix multiplications, each of which represents a new hop. Every iteration has the possibility of stringing together and consolidating more vehicles into a partition. Once a matrix multiplication results in a connectivity matrix identical to the operands, then the connectivity matrix is considered finalized. Any two vehicles $i$ and $j$ which have a $1$ at $a_{i,j}$ are said to be connected together by multi-hop communication. They share the same connectivity partition.

% === II. Application ========================
% =================================================================================

\section{Application}
\subsection{Concept and Purpose}

The purpose of the application is to ingest BSM data and use it to display a simulation of vehicle activity and connectivity with respect to time. The simulator takes as input Basic Safety Messages from multiple vehicles. Ideally, these are BSMs from vehicles that were concurrently producing BSM output and that therefore have similar timestamps. Once these BSMs are uploaded as a CSV, for any given timestamp the simulator can display the position of each vehicle as a pin on a map. The user can progress from one timestamp to the next, watching the pins move along the roads on the map (see Figure \ref{low}).

\begin{figure}[h!]
\centering
\includegraphics[width=3in]{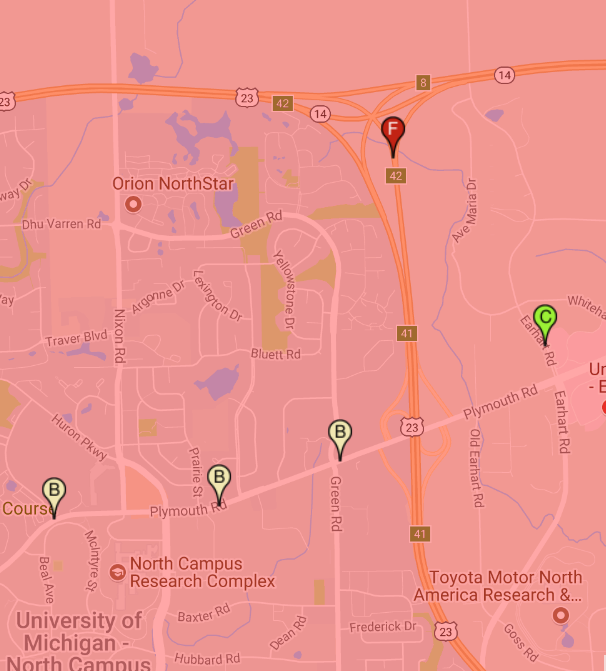}
\caption{Vehicle locations represented using markers of different colors}
\label{pins}
\end{figure}

In addition to its position, each pin has a color and a character(s). Pins with matching color-character combinations are in the same connectivity partition--they are able to send data to each other via multi-hop communication (see Figure \ref{pins}). For example, if two pins are within the DSRC range of each other, they can communicate with a single hop (and will therefore share the same color and character for each timeframe in which they are within range). If two pins are not within the DSRC range, then they cannot communicate unless there exists a middle pin that they can both reach via DSRC. If there is at least one mutually-reachable middle pin between the two pins, the external pins can use the middle pin to help broadcast their message, creating an ad-hoc network and successfully utilizing multi-hop communication. 

In Figure \ref{pins}, vehicle locations are represented with pins. The three vehicles at the bottom left share the same connectivity partition; they have the same color and character because they can reach each other via multi-hop communication. With the passing of each timeframe, the colors and characters on the pins will change. This is because vehicles move in and out of range, alternately leaving and mingling among connectivity partitions. The range of DSRC devices--which determines the maximum distance of a single hop--can be adjusted by the user; the default DSRC range, which is based on prior research, is 1000 meters.

The simulation was created with C\# and ASP.NET; Javascript calls were made to the Google Maps API.

\subsection{Usage}

\begin{figure}[h!]
\centering
\includegraphics[width=3in]{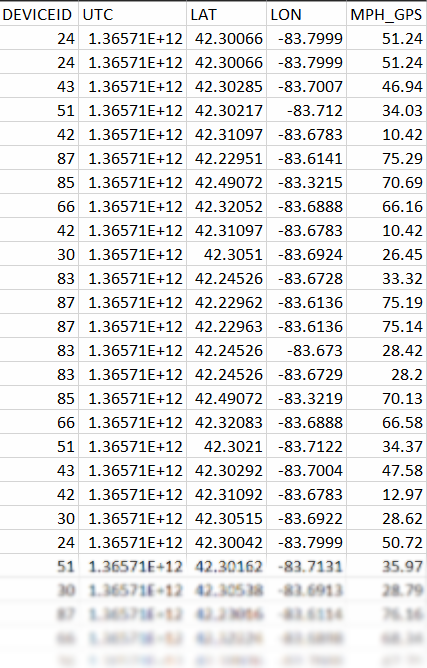}
\caption{Expected CSV input format for the simulation}
\label{expected_input_format}
\end{figure}

An example of the expected input format is a large CSV with the structure of Figure \ref{expected_input_format}. Each row should be an abbreviated form of a BSM; all BSM fields are removed except the five listed. After uploading, the user must only click "Run Simulation" or step through the timestamps one-by-one.

% === II. Performance Analysis ========================
% =================================================================================
\section{Performance Analysis}

A performance analysis of the simulation was conducted. This was done in order to study the efficiency of both the simulation as a whole and the underlying multi-hop partitioning algorithm as the number of vehicles increased within a confined space. The result was a greater understanding of some of the application's performance limitations. The multihop partitioning algorithm was identified as taking the bulk of the computation time. Future performance analysis should study the effects of keeping vehicle density constant as vehicle count changes.

\begin{figure*}[!htb]
\centering
\includegraphics[width=6.5in]{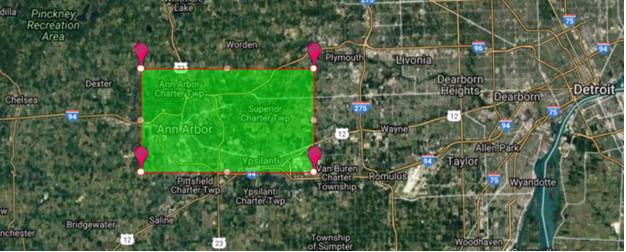}
\caption{Geographical location of the sample dataset used for simulation}
\label{rectangle}
\end{figure*}

\subsection{Process}

The performance testing was done by artificially generating BSM data files and measuring the simulation's performance in milliseconds when run on these files. Each generated file had a different number of vehicles per timestamp ($N$); some had as few as 1 vehicle and others had as many as 200 vehicles per timestamp. All vehicles' positions were constrained within a fixed rectangular area of 348.16 km² in Ann Arbor, Michigan. They would appear at random points within the green rectangle in Figure \ref{rectangle}. Rather than only appearing on road surfaces, they could appear at random anywhere within this rectangle. Also, their positions in one timestamp did not influence their positions in the next timestamp. Whereas vehicles were confined to a given territory in the generated data, as the number of vehicles increased the density rose quickly. 

For test files that contained many vehicles, fewer timestamps were included. This was done to normalize the test file sizes to less than 5000 KB. It was observed during the experiment that the application incurred a significant amount of delay when uploading files larger than about 4500 KB.

The GPS coordinates of the confining rectangle for generated data are as follows, as seen in Figure \ref{rectangle}: (42.356186, -83.522030), (42.356186, -83.816270), (42.226673, -83.522030), and (42.226673, -83.816270).

\begin{figure*}[!htb]
\centering
\includegraphics[width=5.8in]{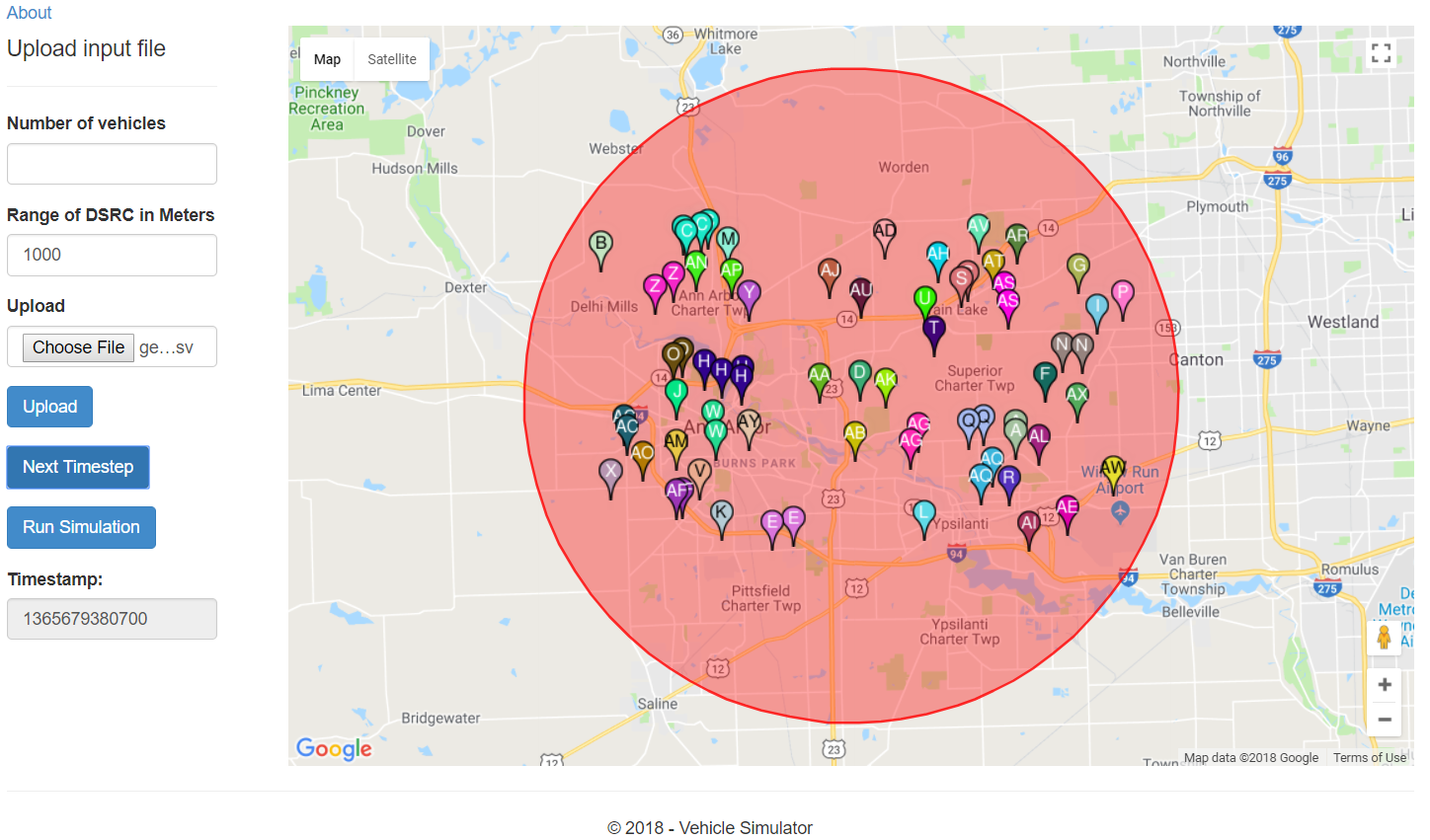}
\caption{Graphical interface of the simulator with sparse distribution of nodes}
\label{low}
\end{figure*}

\begin{figure*}[!htb]
\centering
\includegraphics[width=5.8in]{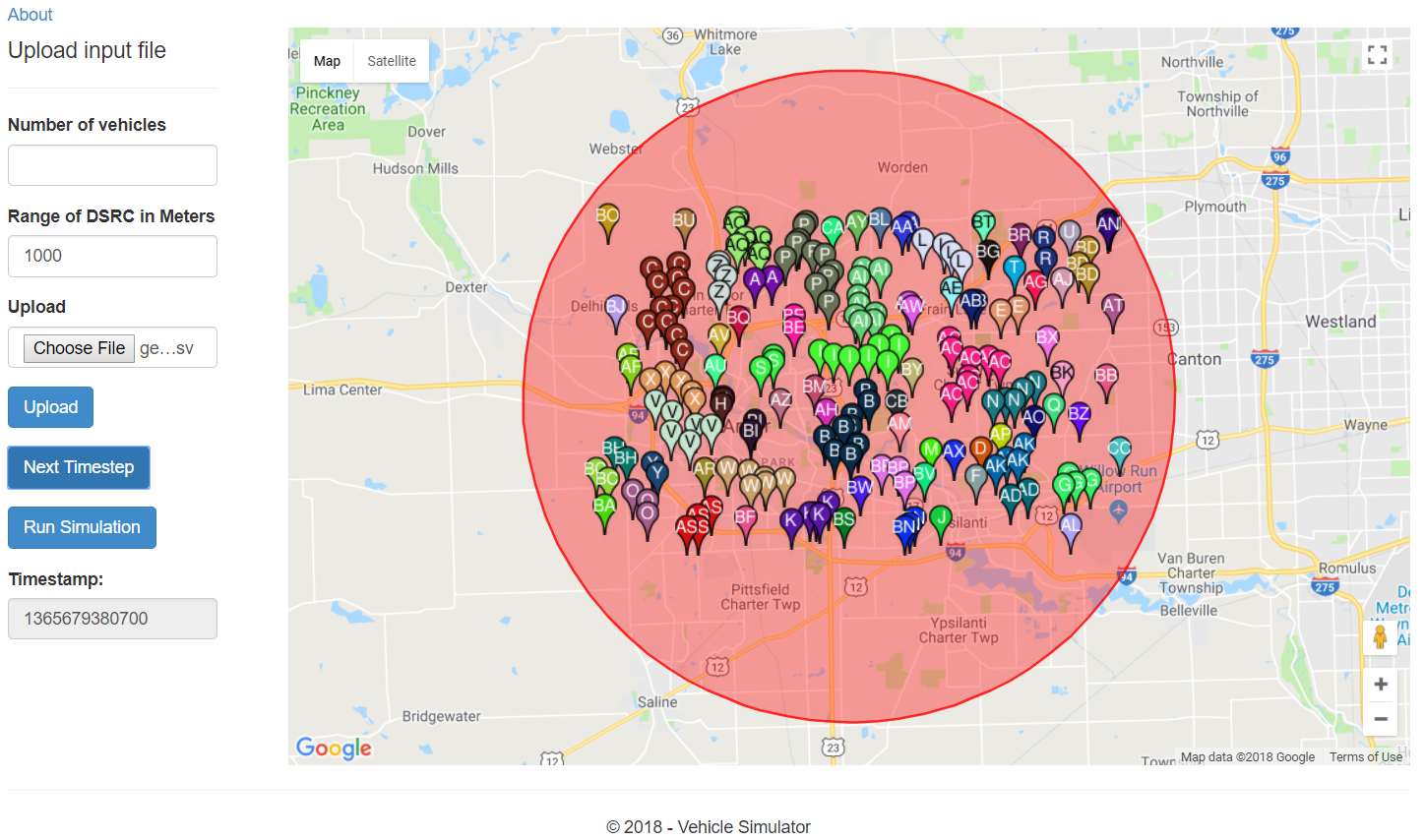}
\caption{Graphical interface of the simulator with high density of nodes}
\label{high}
\end{figure*}

In Figure \ref{low}, the graphical interface for the simulator has few vehicles (it has a sparse distribution). In Figure \ref{high}, however, there is a dense distribution of vehicles and the corresponding partitioning is shown. In Figure \ref{high}, the time to calculate the connectivity matrix has become very noticeable following the rise in vehicle count. The number of partitions has increased to its peak and will begin to lower if additional vehicles are introduced.

\subsection{Results and Conclusions}

\begin{figure}[!htb]
\centering
\includegraphics[width=0.5\textwidth]{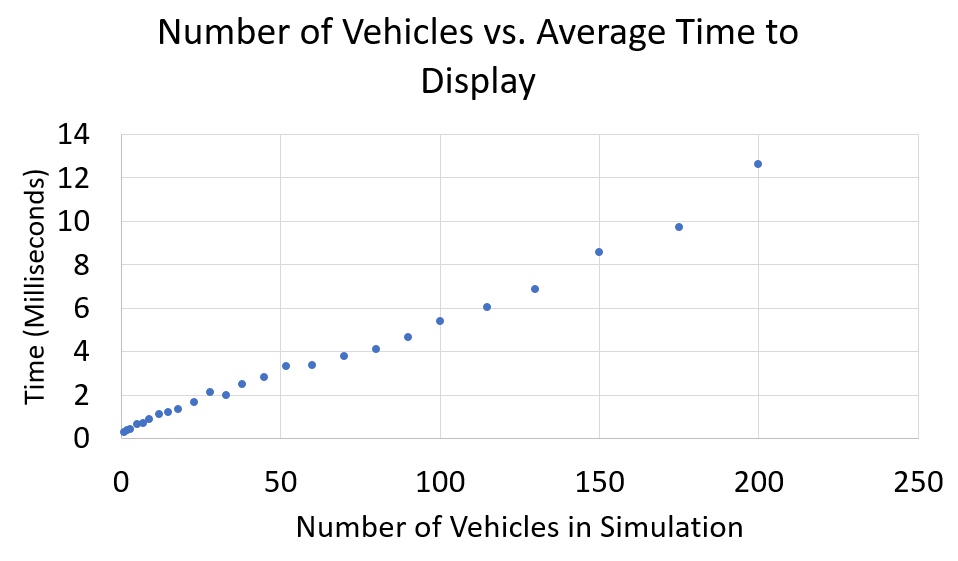}
\caption{Average time to render graphic display}
\label{displayTrend}
\end{figure}

\begin{figure}
\centering
\includegraphics[width=0.5\textwidth]{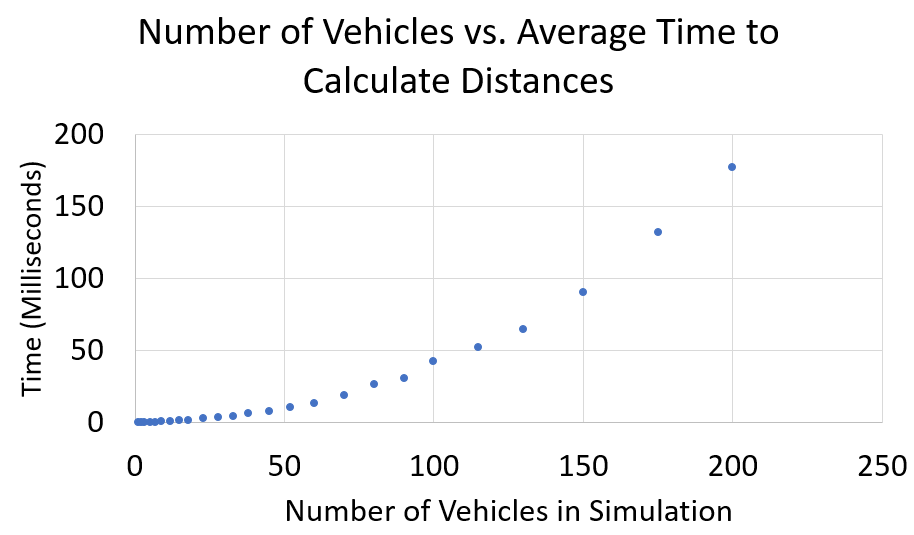}
\caption{Average time to calculate distances between vehicles}
\label{distanceTrend}
\end{figure}

\begin{figure}
\centering
\includegraphics[width=0.5\textwidth]{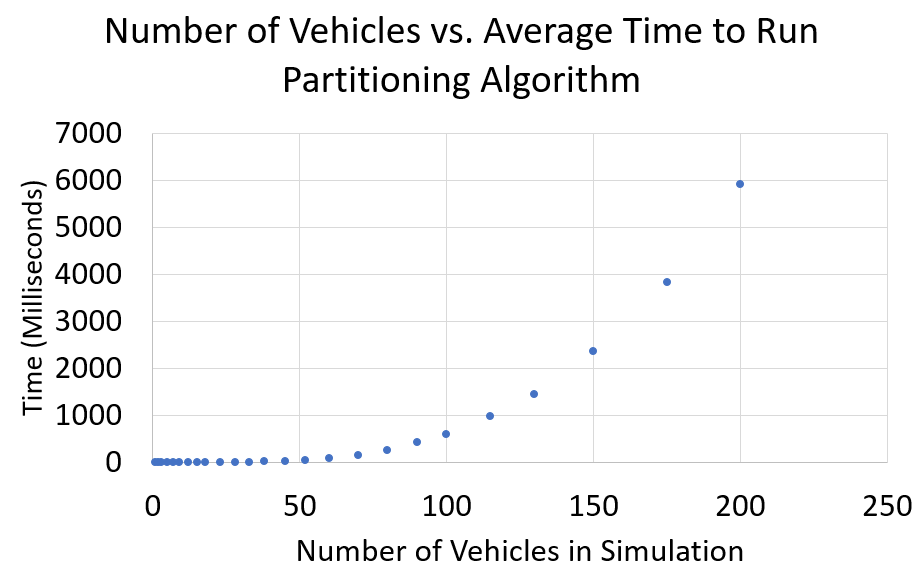}
\caption{Time complexity of partitioning algorithm}
\label{multihopTrend}
\end{figure}

\begin{figure}
\centering
\includegraphics[width=0.5\textwidth]{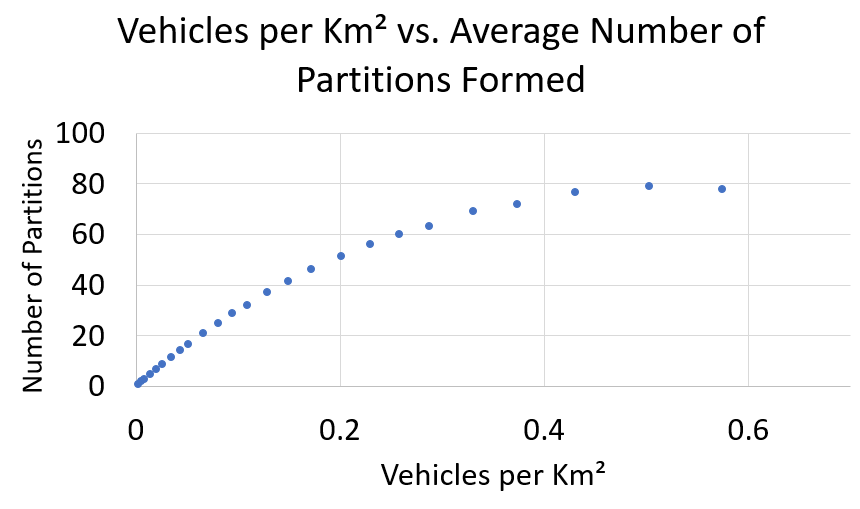}
\caption{Average number of partitions}
\label{partitionsTrend}
\end{figure}

The completion time was measured in milliseconds and logged for various parts of the simulation process. Also logged was the number of individual input vehicles within the BSM timestamp. After numerous runs with different input CSVs, the log files were combed for the data and patterns were observed; see Figures \ref{displayTrend}, \ref{distanceTrend}, \ref{multihopTrend}, and \ref{partitionsTrend}.

The time needed to populate the display with pins was insignificant, even when the population of vehicles was large (see Figure \ref{displayTrend}). This was important to verify because it was necessary to determine whether delays were being caused by the internal multi-hop process or by the display medium.

Note that the simulation began to take considerable time at a certain point (see Figure \ref{multihopTrend}). This may present a problem for those that wish to use the current tool to study very large vehicle populations (in excess of 200 vehicles). In the future, an `export simulation' functionality could be implemented that could allow the user to run the simulation in the background and output the results to a file, to be displayed later.

From Figure \ref{displayTrend}, it appears that time required for rendering the graphic display would increase linearly as the number of vehicles increases. The display of each additional vehicle requires only the fetching of the appropriate pin image from Google's server and the display of that pin. Ultimately, time to display was insignificant (relative to the amount of time needed to perform the consecutive matrix multiplications to form the multi-hop partitions). Note however that this depends on the strength of Internet connection.

In Figure \ref{distanceTrend}, the distance calculation has become more expensive as the number of vehicles has increased. A prerequisite to performing the partition calculations was the calculation of the distance from every vehicle to every other vehicle. These calculations needed to take place exactly once before the partitions of any particular timestamp could be determined. Calculating these distances involved approximately $n^2$ amount of work. (The formula for this is $n(n-1)/2$, which is close to $n^2$.) The results were in accordance with this expectation. Though the calculation of distances would have eventually become a problem as the number of vehicles increased, it would still be a tiny fraction of the amount of work needed to perform the partition calculations (assuming that the vehicle density was allowed to increase).

A series of consecutive matrix multiplications forms the multi-hop partitions. In Figure \ref{multihopTrend}, we see that as the number of vehicles increased, the multi-hop algorithm became the most costly part of the simulation. It was concluded that this algorithm is overall the most work-intensive part of the simulation by far. At 200 vehicles (0.5 vehicles per km$^2$ ), each timestamp was taking 6 seconds on average.

The simulation was run each time assuming that each vehicle had a DSRC range of 1000 meters (1 km). As the number of vehicles within the rectangle increased, so did the number of partitions (Figure \ref{partitionsTrend}). The number of partitions increased as vehicle density increased, up to a saturation point. However, the rate of increase declined as the density grew. It was observed that at a density of 0.5 vehicles per km$^2$, the number of partitions began to decrease. If even greater densities had been tested, the number of partitions would have fallen to 1. It is expected that at a density of 1 vehicle per km$^2$ (where the square root of the inverse of the density equals the transmission range) there would have been 1 partition on average. Again, however, the artificial data allowed vehicles to disperse themselves randomly. In real data, vehicles would limit themselves to the roads, effectively reducing the space between vehicles. Thus for real data the density at which there would have been only a single partition would have been considerably less than that of this artificial data. This is especially true when the roads are less like a grid and more like a highway, or when the DSRC transmission range is increased by flat land and few obstacles.

\subsection{Limitations}

The program written to artificially generate BSM data permitted vehicles to be randomly generated anywhere in the rectangular area. In reality, vehicles stay on roads and often share the same road. They do not have the freedom of going everywhere. Thus, in reality, vehicles will be less sparse and less dispersed than they are in the artificially-generated data. This means that, if real BSM data had been used for performance analysis, there likely would have been more partitions than were observed. The number of partitions could have thereby had an impact on the time to calculate the partitions. There is therefore a certain degree of uncertainty regarding the extent to which the simulation will perform similarly for real data. Since this application is intended to assist transportation researchers and analysts, who generally focus on real BSM data, in the future we will run the performance analyses with real data.

% === II. Conclusion ========================
% =================================================================================
\section{Conclusion}
An intuitive visual simulator has been developed to assist transportation researchers and analysts to study BSM data. This paper has documented the application and shared the details of its performance analysis. The application successfully generates time-ordered displays that represent vehicle positions and connectivity statuses. The complexity of the partitioning algorithm has been determined to become the a bottleneck when the number of vehicles increases within a confined space. In a future build of the tests, it will be a goal to compare the effects of vehicle population size with those of vehicle population density. For example, keeping vehicle density constant while increasing vehicle population may produce interesting results. In addition, another advancement would be to perform the performance analyses with real data instead of generated data.

% ====== REFERENCE SECTION

%\begin{thebibliography}{1}

% IEEEabrv,

\bibliographystyle{IEEEtran}
%\bibliography{IEEEabrv,Bibliography}
\bibliography{Bibliography.bib}

%\end{thebibliography}
% biography section

% that's all folks
\end{document}